\documentclass[aps,twocolumn,showpacs,preprintnumbers,amsmath,amssymb,floats,citeautoscript,nobalancelastpage]{revtex4-1}
\usepackage{graphicx}
\usepackage{dcolumn}
\usepackage{bm}
\usepackage{txfonts}
\usepackage{multirow}
\usepackage{color}
 
\begin{document}

\title{Anomalous Thermoelectric Response in an Orbital-Ordered Oxide Near and Far from Equilibrium}

\author{Y.~Nishina$^{1}$}
\author{R.~Okazaki$^{2,\ast}$}
\author{Y.~Yasui$^{3}$}
\author{F.~Nakamura$^{4}$}
\author{I.~Terasaki$^{1}$}

\affiliation{$^1$Department of Physics, Nagoya University, Nagoya 464-8602, Japan}
\affiliation{$^2$Department of Physics, Faculty of Science and Technology, Tokyo University of Science, Noda 278-8510, Japan}
\affiliation{$^3$Department of Physics, Meiji University, Kawasaki 214-8571, Japan}
\affiliation{$^4$Department of Education and Creation Engineering, Kurume Institute of Technology, Kurume 830-0052, Japan}

\begin{abstract}
We report the thermoelectric transport properties in the orbital-ordered Mott insulating phase of Ca$_2$RuO$_4$
close to and far from equilibrium.
Near equilibrium conditions where the temperature gradient is only applied to the sample, 
an insulating but non-monotonic temperature variation of the Seebeck coefficient is observed, 
which is accounted for in terms of a temperature-induced suppression of the orbital order.
In non-equilibrium conditions where we have applied high electrical currents,
we find 
that the Seebeck coefficient is anomalously increased in magnitude with 
increasing external current.
The present result clearly demonstrates a non-thermal effect
since the heating simply causes a decrease of the Seebeck coefficient,
implying a non-trivial non-equilibrium effect
such as a modification of the spin and orbital state in currents.
\end{abstract}

\maketitle

Understanding various phenomena arising from
interplayed spin and orbital degrees of freedom in correlated electrons,
such as high-$T_c$ superconductivity and colossal magnetoresistance, 
is a central topic in contemporary condensed matter physics \cite{Dagotto05}.
Recently, it also becomes a challenging issue 
to explore
how such an electronic state is changed in non-equilibrium situations \cite{Yonemitsu2008,Aoki2014},
which are realized under laser irradiations \cite{Koshihara2006}
or by applying high electric fields \cite{Asamitsu1997}.
Intriguingly, the photo-induced or field-induced state
might be essentially distinguished from thermally-excited one \cite{Potember1979,Tayagaki2001,Sawano2005,Kim2008},
stimulating further investigation for an inherent non-equilibrium property of correlated electrons.

The Mott insulator Ca$_{2}$RuO$_4$ \cite{Nakatsuji1997,Cao1997} 
offers a unique playground to study the non-equilibrium effect
on correlated electron systems.
In equilibrium, this compound exhibits 
a first-order metal-insulator transition
at $T_{\rm MI} \simeq$ 360~K \cite{Alexander1999}
and an antiferromagnetic transition at 
$T_N = 110$ K.
The optical conductivity \cite{Jung2003},
resonant x-ray \cite{Kubota2005} and
perturbed angular correlation \cite{Rams2009}
experiments have revealed that 
a ferro-type $d_{xy}$ orbital ordering is realized
below $T_{\rm MI}$, 
as is expected from flattening of the RuO$_6$ octahedra at low temperatures \cite{Braden1998,Fridet2001,Anisimov2002,Gorelov2010,Sutter2017}, 
while different orbital patterns are also suggested \cite{Mizokawa2001,Hotta2001,Lee2002}.
Recently, Nakamura {\it et al.} reported that this compound exhibits 
an electric-field-induced insulator-to-metal transition at
$E_{\rm th} \sim$ 40 V/cm at room temperature \cite{Nakamura2013},
which is very low threshold field compared with that for Zener and avalanche breakdown \cite{Conwellreview}.
The subsequent experiment using a non-contact infrared thermometer has revealed that,
although it does not reach the transition point,
the nonlinear conduction indeed occurs in the isothermal environments below $E_{\rm th}$ \cite{Okazaki2013},
indicating a novel non-equilibrium state in this system.

In this paper, 
we investigate how the non-equilibrium parameter affects 
the orbital ordering phase of Ca$_2$RuO$_4$
by means of thermoelectric transport measurement near and far from equilibrium.
The Seebeck coefficient studied here is a powerful probe to clarify
the electronic state \cite{Mahan1998,Behnia2004}
including the spin and orbital nature of correlated electrons \cite{Koshibae2000,Koshibae2001}. 
First we show the Seebeck coefficient measured without applying 
external currents. 
The temperature variation is found to be insulating but non-monotonic, 
which is qualitatively explained by the suppression of the orbital ordering with heating.
In non-equilibrium, we have measured the Seebeck coefficient with applying currents,
and find that the absolute value of the Seebeck coefficient is surprisingly increased with increasing currents.
Since the heating leads to a decrease of the Seebeck coefficient,
the present result clearly shows an inherent  non-equilibrium effect,
which we suggest to be attributed to a current-induced quenching 
of the spin and orbital degrees of freedom of the Ru $4d$ electrons.

The experiments were performed using 
Ca$_2$RuO$_4$ single crystals with typical sample dimensions of $2\times2\times0.2$\,mm$^3$ grown by a floating-zone method \cite{Nakatsuji2001}. 
Near equilibrium condition,
the Seebeck coefficient was measured with a conventional steady-state method 
using a copper-constantan differential thermocouple
from 360 down to 180 K in an electrical furnace and a liquid helium cryostat.
In the present experiments, 
the sample temperature was kept below $T_{\rm MI}$
because it is highly probable that the crystal will be broken 
owing to the large structural change at $T_{\rm MI}$ \cite{Fridet2001}.
The experimental setup in the non-equilibrium condition is 
schematically shown in Fig.~1(a).
The temperature gradient $\nabla T$~($\parallel ab$~planes) is controlled by using two Peltier modules,
the gap of which was bridged with a sapphire plate.
The sample is fixed on the sapphire using varnish and its temperature gradient is 
measured by a non-contact infrared thermography (InfReC R300, Nippon Avionics).
This method is highly advantageous to directly measure the temperature 
without additional heat capacity and contact thermal resistance
of conventional contact-type thermometers \cite{Okazaki2013}. 
The four electrical contacts were carefully made with a gold deposition technique to reduce the contact resistance.

\begin{figure}[t]
\begin{center}
\includegraphics[width=8cm]{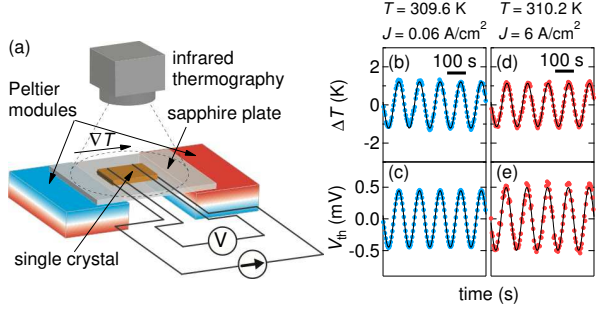}
\caption{(Color online).
(a) Schematic figure of the experimental configuration
for measurements of the Seebeck effect in non-equilibrium.
The temperature gradient $\nabla T$ is controlled by using two Peltier modules,
which are bridged with a sapphire plate.
The single crystal is fixed on the sapphire and the temperature gradient is directly
measured by an infrared thermography.
The right panels display the
time variations of (b,d) the temperature difference between the voltage contacts $\Delta T$ and 
(c,e) the thermoelectric voltage $V_{\rm th}$
for (b,c) low and (d,e) high current densities.
The solid curves show the sinusoidal fitting results.
}
\end{center}
\end{figure}

In the present thermoelectric measurement in the external current $I$, the measured voltage $V$  is expressed as 
\begin{equation}
V = V_R+V_{\rm th}+V_0 = R(I)I+S(I)\Delta T +V_0,
\end{equation}
where $R(I)$ and $S(I)$ are current-dependent resistance and Seebeck coefficient, respectively,
$\Delta T$ is the temperature difference between the voltage contacts,
and $V_0$ is an extrinsic offset term.
Since the resistive voltage $V_R$ is 
much larger than the thermoelectric one $V_{\rm th}$ in currents,
we used a double ac method as follows: 
By applying ac current with a rectangular waveform (2.5 Hz),
we obtain 
$R =V_R/I = (V_+-V_-)/2I$
and 
$V_{\rm th}+V_0 = (V_++V_-)/2$,
where $V_+(V_-)$ is the measured voltage with applying current for positive (negative) direction.
We further vary the temperature difference with a slow sinusoidal waveform (0.01 Hz)
as shown in Fig. 1(b,d).
Correspondingly the thermoelectric voltage oscillates with the same waveform as seen in Fig. 1(c,e),
in which the offset contribution is subtracted,
and then $S(I)$ is  calculated from the amplitudes in these oscillations.

\begin{figure}[t]
\begin{center}
\includegraphics[width=8cm]{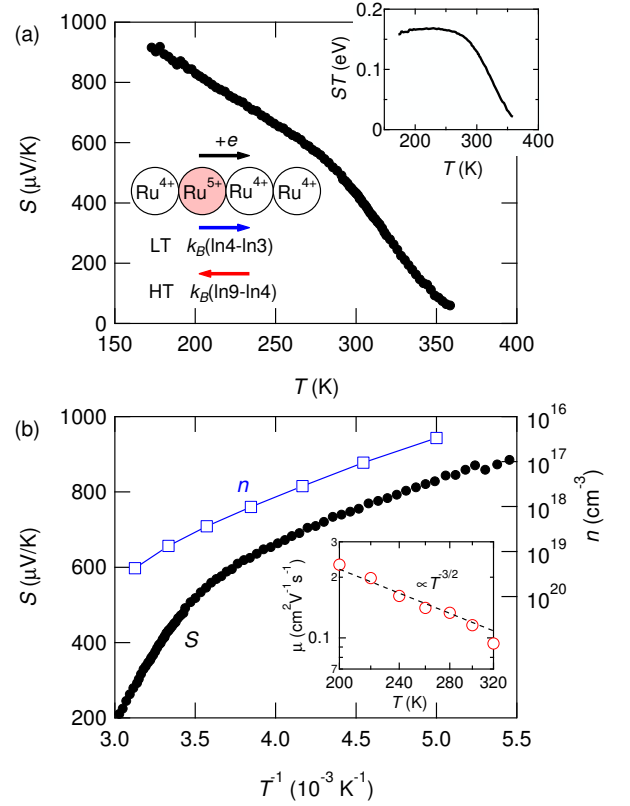}
\caption{(Color online).
(a) The Seebeck coefficient $S$ measured with applying no electrical current as a function of temperature $T$.
The upper right inset shows $ST$ as a function of $T$.
The lower left inset is a schematic illustration of the carrier transport. 
At low temperatures the hole carries a positive entropy of $k_B\ln(4/3)$, while 
it flows with a negative entropy of $k_B\ln(4/9)$ at high temperatures.
For details see text.
(b) $S$ (black circles, left axis) and 
the carrier concentration $n$ (blue squares, right axis) as a function of $T^{-1}$.
The inset is a log-log plot for the temperature variation of mobility $\mu$.
The dashed line represents a $T^{-3/2}$ dependence.
}
\end{center}
\end{figure}

We first discuss the thermoelectric transport near equilibrium.
Figure~2(a) depicts the temperature dependence of the Seebeck coefficient measured with 
applying no external current.
The Seebeck coefficient with positive sign decreases in magnitude with increasing temperature.
We find that, above around 270 K,
the Seebeck coefficient is more strongly suppressed than that expected from 
the non-degenerate formula $S(T) = \frac{k_B}{e}\frac{\Delta}{k_BT}$,
where $k_B$ is the Boltzmann constant, $e$ is the elementary charge, and 
$\Delta$ is an energy gap measured from the valence-band top to the Fermi level.
This anomaly is clearly seen in a plot of $ST$ as a function of $T$
shown in the upper right inset of Fig. 2(a),
in which $ST$ deviates from a constant value above 270 K.

To clarify an origin of the anomalous suppression of the Seebeck coefficient toward $T_{\rm MI}$,
we have measured the Hall effect 
because the Seebeck coefficient is often discussed in terms of the carrier concentration \cite{Mahan1998}.
The Hall resistivity $\rho_{yx}$ exhibits a linear $H$ dependence up to $\mu_0H = 7$ T (not shown),
and we obtain the carrier concentration 
$n = e^{-1}R_H^{-1}$.
Note that the Hall coefficient $R_H=d\rho_{yx}/d(\mu_0H)$ is positive, 
as is indicated from the positive Seebeck coefficient shown in Fig.~2(a).
Here we plot $S$ and $\ln n$ as a function of $T^{-1}$ in Fig.~2(b)
for comparison.
In non-degenerate semiconductors, the Seebeck coefficient is related to $n$ as 
$S(T) = \frac{k_B}{e}\frac{\Delta}{k_BT} =\frac{k_B}{e}\ln(\frac{N_v}{n})$,
where $N_v$ is the effective density of states in the valence band.
In the present case, 
while $S(T)$ is drastically suppressed and deviates from the $T^{-1}$ dependence above around 270 K,
$\ln n(T)$  varies almost linearly against $T^{-1}$ even above 270 K.
Thus this result shows that 
the anomalous temperature variation of the Seebeck coefficient does not stem from
temperature variation of the carrier concentration.
We note that the optical gap varies linearly with temperature as 
$\Delta(T) = \Delta_0+aT$, where $\Delta_0$ and $a$ are temperature-independent constants,
in the present temperature range \cite{Jung2003}.
Also, this is not the origin because in this case
the formula is rewritten as 
$S(T) = \frac{k_B}{e}\frac{\Delta_0}{k_BT} + \frac{a}{e}$, 
which still shows a $T^{-1}$ dependence.

Here we focus on the spin and orbital state in the orbital ordering phase,
which affects the Seebeck coefficient through its degeneracy
as discussed in several transition-metal oxides \cite{Terasaki1997,Palstra1997,Kobayashi2004}.
In a temperature range of $t\ll k_BT \ll U$, 
where $t$ is the transfer energy (bandwidth) and $U$ is the Coulomb interaction,
the Seebeck coefficient is expressed as 
\begin{equation}
S = -\frac{k_B}{e}\ln\left(\frac{g_{4+}}{g_{5+}}\frac{x}{1-x}\right),
\label{koshibaeeq}
\end{equation}
known as an extended Heikes formula \cite{Koshibae2000}.
Here $g_{4+}$ and $g_{5+}$ are the degeneracy of the electron configuration
of Ru$^{4+}$ and Ru$^{5+}$ sites, respectively, and 
$x = M/N_A$ ($M$ is the number of Ru$^{5+}$ sites and $N_A$ is the total number of Ru sites)
is the fraction of the Ru$^{5+}$ ions, 
equal to the hole concentration per Ru site.
Note that
the measured temperature range of $T_N < T < T_{\rm MI}$ may satisfy the temperature range valid for Eq. (\ref{koshibaeeq}),
since $T_N$ can be scaled to $t^2/U$ and $T_{\rm MI}$ to $U$.
Also suggested that the Heikes formula still holds at low temperatures where $k_BT < t$ \cite{Palsson1998}.
Now $n = 10^{19}$ cm$^{-3}$ at $T \sim 310$ K
corresponds to 
$x \simeq 1\times 10^{-3}$ holes per Ru atom,
which is much less than unity. Thus Eq. (\ref{koshibaeeq}) is reduced to
\begin{equation}
S = -\frac{k_B}{e}\ln x -\frac{k_B}{e}\ln\left(\frac{g_{4+}}{g_{5+}}\right).
\label{koshibaeeqa}
\end{equation}
The first term is positive and related to the hole concentration $n$.
The second term, which shows the ratio of the entropy carried by a carrier to the charge,
can take either positive or negative, depending on the degeneracy in the Ru sites.

\begin{table}[b]
\caption{\label{table1} 
Degeneracy of the electron configuration of the Ru sites.}
\begin{ruledtabular}
\begin{tabular}{cccc}
  & Ru$^{4+}$ (LT) & Ru$^{4+}$ (HT) & Ru$^{5+}$ \\
\hline
$g_{\rm spin}$  & 3 & 3 & 4\\
$g_{\rm orbital}$ & 1 & 3 & 1\\
$g_{\rm total}$ & 3 & 9 & 4\\
\end{tabular}
\end{ruledtabular}
\end{table}

The degeneracy of the Ru $4d$ electrons is
summarized in Table I.
Four and three $4d$ electrons in the $t_{2g}$ manifolds 
lead to the spin magnitude (the spin degeneracy) of 1 ($g_{\rm spin}=3$) and 3/2 ($g_{\rm spin}=4$)
for Ru$^{4+}$ and Ru$^{5+}$ sites, respectively.
Then we consider the the orbital degeneracy $g_{\rm orbital}$.
For Ru$^{4+}$ sites, at low temperatures (noted as LT in Table. I), 
the $d_{xy}$ orbital is well ordered, 
meaning that the orbital degeneracy is completely quenched ($g_{\rm orbital} = 1$).
At high temperatures (HT), on the other hand, the $d_{xy}$ orbital polarization is 
decreased with heating.
This means that the orbital degeneracy {\it recovers} with increasing temperature toward $T_{\rm MI}$
and approaches to $g_{\rm orbital} = 3$.
Consequently, for Ru$^{4+}$ sites, the total degeneracy $g_{\rm total}=g_{\rm spin}g_{\rm orbital}$
increases from 3 to 9 with heating.
For Ru$^{5+}$ sites, $g_{\rm orbital}$ is always unity
since three electrons occupy the lower Hubbard $t_{2g}$ manifolds \cite{Anisimov2002} and
then $g_{\rm total} = 4$ holds in the entire temperature range.
The degeneracy ratio of $g_{4+}/g_{5+}$ thus varies
from $3/4$ to $9/4$ with heating, and
therefore the degeneracy term in Eq. (\ref{koshibaeeqa}) 
positively contributes to the Seebeck coefficient at low temperatures but 
negatively at high temperatures close to $T_{\rm MI}$.
This gives a qualitative explanation for the observed suppression of the Seebeck coefficient 
toward $T_{\rm MI}$.
Such a hole transport is schematically illustrated in the lower left inset of Fig.~2(a).
It should be noted that 
the interference intensity in the resonant x-ray scattering technique,
which represents the degree of $d_{xy}$ orbital polarization,
is reduced above around 220 K \cite{Kubota2005}.
This is close to the temperature above which the Seebeck coefficient 
deviates from the $T^{-1}$ dependence in the present study.

\begin{figure}[b]
\begin{center}
\includegraphics[width=7cm]{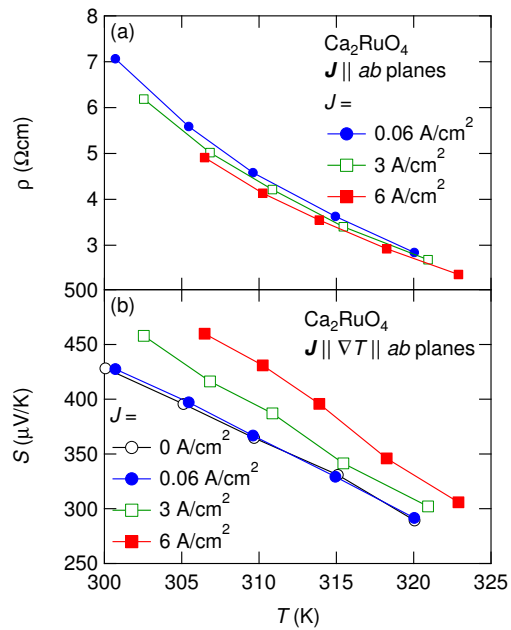}
\caption{(Color online).
Temperature dependence of (a) the electrical resistivity $\rho$ 
and (b) the Seebeck coefficient $S$ measured with applying several current densities $J$. 
}
\end{center}
\end{figure}

We now examine the non-equilibrium current effect on the Seebeck coefficient.
Figures~3(a) and 3(b) show temperature variations of the resistivity
and the Seebeck coefficient measured with applying current density $J$ using the setup shown in Fig.~1(a). 
The resistivity indeed decreases with increasing currents in the isothermal conditions, 
which has been suggested to be attributed to a current-induced energy-gap suppression \cite{Okazaki2013}.
There, the current acts as an injection of excess quasiparticles,
as proposed in non-equilibrium superconducting state \cite{Owen1972} and charge ordering \cite{Ajisaka2009}.
Note that the present field range is too small to drive the conventional Zener or avalanche breakdown phenomena \cite{Nakamura2013,Okazaki2013}.
Hot electron phenomenon, in which 
the electronic temperature $T_e\equiv\varepsilon_{\rm kin}/k_B$ ($\varepsilon_{\rm kin}$ is the kinetic energy of electron)
becomes larger than the lattice temperature $T_l$ to produce the nonlinearity \cite{Conwellreview},
is also unlikely.
In this compound, as shown in the inset of Fig.~2(b), the mobility $\mu=(ne\rho)^{-1}$ exhibits a $T^{-3/2}$ dependence,
indicating that the acoustic phonon scattering is dominant.
In this regime, the relaxation time depends on the energy as $\tau(\varepsilon)\propto \varepsilon^{-1/2}$.
Thus an increase of $T_e$ leads to a decrease of $\tau$, and
the conductivity decreases with increasing currents as observed in Si and Ge \cite{Ryder1953}.
This model does not adapt the present result since the conductivity increases with currents in Ca$_2$RuO$_4$.

The most highlighted result is the current effect on the Seebeck coefficient shown in Fig.~3(b), because 
the Seebeck coefficient increases with currents, in total contrast to the thermal effect that reduces its magnitude.
This is also distinguished from the results in other systems that exhibit the nonlinear conduction.
In the charge-density-wave (CDW) materials, the sliding motion of depinned CDW causes a nonlinear conduction in fields \cite{Gruner1988}.
There, both the resistivity and the Seebeck coefficient decrease with increasing fields \cite{Stokes1984}, 
in high contrast to the present study.
Hot electron characterized by its large kinetic energy may enhance the Seebeck coefficient \cite{Zebarjadi2007},
since the Seebeck coefficient measures the energy difference 
from the Fermi level to the electron energy as seen in its non-degenerate formula.
The current-induced change in the Seebeck coefficient is given as
$\Delta S \sim \frac{mJ^2}{e^3n^2T}$ \cite{Zebarjadi2007},
where $m$ is the electron mass, yielding 
$\Delta S \sim 10^{-16}$~V/K for Ca$_2$RuO$_4$ at $T=300$~K and $J = 6$~A/cm$^2$.
This estimated value is much smaller than the change of $\sim 100$ $\mu$V/K observed in the present study,
again indicating that hot electron is not important in this system.

The present result of the Seebeck effect seems to be incompatible with the current-induced gap suppression
suggested from the nonlinear resistivity measurement \cite{Okazaki2013},
since the gap reduction usually leads to the decrease of the Seebeck coefficient.
Instead, we again consider the degeneracy of the Ru $4d$ electrons
on the basis of the extended Heikes formula of Eq. (\ref{koshibaeeqa}).
The first term is decreased when the energy gap is suppressed by currents,
while the reduction is small due to the weak nonlinear conduction in this current range.
We then focus on the second term.
Now applying external current makes the system more metallic, indicating that
the localized model for counting such a degeneracy becomes less effective with increasing currents.
Therefore, in currents, 
the correlated hole with the degeneracy $g_{\rm 5+} = 4$ 
is regarded to flow
among the itinerant Ru sites in which the spin and orbital degeneracy is quenched owing to 
the current-induced metallization.
This picture is contrast to the conduction near equilibrium
in which the hole flows among the localized Ru sites.
In the itinerant model, the second term in Eq. (\ref{koshibaeeqa}) becomes a positive value of
$+\frac{k_B}{e}\ln4$.
This is larger than the values near equilibrium, 
possibly contributing to the enhancement of the Seebeck coefficient far from equilibrium,
while further microscopic investigation for clarifying the non-equilibrium spin and orbital state is required as a future study.

In summary, we report the anomalous thermoelectric transport in the orbital-ordered Mott insulating phase of Ca$_2$RuO$_4$
near and far from equilibrium.
The Seebeck coefficient near equilibrium is qualitatively explained by the degeneracy of the spin and orbital state of Ru $4d$ electrons.
Far  from equilibrium, we find a non-trivial current-induced enhancement of the Seebeck coefficient in isothermal conditions,
which is difficult to be explained by conventional mechanisms such as hot electron.
We suggest a
quenching of the spin and orbital degrees of freedom due to
the current-induced metallization for understanding the observed non-equilibrium phenomena.

We thank S. Ajisaka, Y. Maeno, S. Nakamura, Y. Nogami, T. Oka, C. Sow, K. Tanabe, H. Taniguchi, K. Toda for stimulating discussion and 
T. D. Yamamoto, T. Kurematsu, M. Sakaki, Y. Kimura for experimental assistance.
This work was supported by the JSPS KAKENHI (No. JP17H06136, No. 26247060, No. 23740266, No. 26610099).

\end{document}